\documentclass[preprint,prd,noshowpacs,nofootinbib]{revtex4}

\usepackage{color} 
\usepackage{amssymb}
\usepackage{amsmath}
\usepackage{graphicx} 
\usepackage{float}
\usepackage{braket}     
\usepackage{pdfpages}
\usepackage{epstopdf}
\usepackage[utf8]{inputenc}

\begin{document}

\title{Comments on the cosmological constant in generalized uncertainty models}

\author{Michael Bishop}
\email{mibishop@mail.fresnostate.edu}
\affiliation{Mathematics Department, California State University Fresno, Fresno, CA 93740}

\author{Joey Contreras}
\email{mkfetch@mail.fresnostate.edu}
\affiliation{Physics Department, California State University Fresno, Fresno, CA 93740}

\author{Peter Martin}
\email{kotor2@mail.fresnostate.edu}
\affiliation{Physics Department, California State University Fresno, Fresno, CA 93740}

\author{Douglas Singleton}
\email{dougs@mail.fresnostate.edu}
\affiliation{Physics Department, California State University Fresno, Fresno, CA 93740}

\date{\today}

\begin{abstract}
 The existence of a small, non-zero cosmological constant is one of the major puzzles in fundamental physics. Naively, quantum field theory arguments would imply a cosmological constant which is up to 10$^{120}$ times larger than the observed one. It is believed a comprehensive theory of quantum gravity would resolve this enormous mismatch between theory and observation. In this work, we study the ability of generalized uncertainty principle (GUP) models, which are phenomenologically motivated models of quantum gravity, to address the cosmological constant problem. In particular, we focus on how these GUP models may change the phase space of QFT, and how this affects the momentum space integration of the zero-point energies of normal modes of fields. We point out several issues that make it unlikely that GUP models, in their current form, would be able to adequately address the cosmological constant problem. 
\end{abstract}

\maketitle

\section{Introduction}
  
A theory of quantum gravity, although not yet a reality, has been advertised as being able to solve many of the ills of classical general relativity, such as the singularities that occur in black hole and cosmological solutions \cite{penrose,hawking}. Quantum gravity is also supposed to resolve some of the issues surrounding the results of applying quantum field theory in a curved space-time such as what happens to a black hole at the end of evaporating via Hawking radiation \cite{hawking-2}, and what happens to the information stored in a black hole due to this evaporation \cite{susskind}.

The puzzle we address in this work is the apparent mismatch between the observed cosmological constant and the theoretically calculated cosmological constant -- a conundrum known as the cosmological constant problem. This cosmological constant problem has been known for a long time. A nice relatively recent review of the issue is reference \cite{weinberg}. The problem is that having a cosmological constant, $\Lambda$, is equivalent to having a constant energy density, $\rho_{vac}$, as a source in the Einstein field equations. The relationship is (using the units and notation of \cite{weinberg})
\begin{equation}
    \label{rho-cc}
    \rho_{vac} = \frac{\Lambda}{8 \pi G}~.
\end{equation}
The subscript $vac$ comes from quantum field theory where one obtains a constant vacuum energy density by adding up all the energy zero modes of vacuum quantum fields. The zero modes are given by $\frac{1}{2} \hbar \omega _p = \frac{1}{2} E_p = \frac{1}{2} \sqrt{{\vec p}^2 +m^2}$, and summing these up to get the vacuum energy density yields
\begin{equation}
    \label{rho-vac}
    \rho_{vac} = \int \frac{d^3p}{(2\pi)^3} \frac{1}{2} \sqrt{{\vec p}^2 +m^2} = \frac{1}{2} \int _0 ^{p_c} \frac{4 \pi}{(2 \pi)^3} dp ~ p^2 \sqrt{p^2 +m^2} \approx \frac{p_c ^4}{16 \pi^2} ~.
\end{equation}
Note that $p=|{\vec p}|$, we use these notations interchangeably throughout the rest of the paper. 
Since the integral is divergent, we cut off the $dp$ integration at some scale $p_c$, which is usually taken to be the Planck scale, $p_c \sim (8 \pi G)^{-1/2}$. Using  \eqref{rho-vac} gives $\rho _{vac} \approx 2 \times 10^{71}$ GeV$^4$. In contrast, the measured vacuum energy density \cite{pdg} is about $\rho _{vac} \approx 10^{-47}$ GeV$^{4}$. The difference between the theoretically calculated $\rho _{vac}$ from \eqref{rho-vac} versus experimentally measured $\rho_{vac}$ is a difference of 118 orders of magnitude. This massive discrepancy {\it is} the cosmological constant problem. Even if one lowers the cut off scale to the QCD scale of $\Lambda_{QCD} \sim 200$ MeV, where we think we fully understand QFT, one still gets a disagreement between theory and experiment of 41 order of magnitude. Some drastic change in our understanding of either QFT, general relativity, or both is needed to resolve this puzzle. 

\section{Generalized Uncertainty Principle and quantum gravity}

One of the proposed resolutions to the cosmological constant problem is a theory of quantum gravity, a catch-all solution to all open problems in fundamental theoretical physics. In this work, we utilize the phenomenological generalized uncertainty principle (GUP) approach to quantum gravity. The GUP approach to quantum gravity is a bottom up approach (in contrast to the more top down approaches to quantum gravity such as superstring theory \cite{polchinski} or loop quantum gravity \cite{rovelli}). There is a vast amount of literature on GUP, with a few of the important representative papers being \cite{vene,gross,amati2,amati,maggiore,garay,KMM,scardigli,adler-1999,adler-2001}. After this original burst of work on GUP there were various other works, a sample of where can be found in references \cite{myung,tzhu,das,nicolini,das-2,brito,ali,faizal} which further developed this area of research. There are also some very recent works \cite{fadel,tamburini} which deal with the algebraic and physical structure of spacetime in connection with GUP. 

The basic idea is that quantum gravity should modify the standard position and momentum commutator of canonical quantum mechanics from $[{\hat x_i}, {\hat p}_j] = i \delta _{ij} \hbar$ to $[{\hat X}_i, {\hat P}_j] = i \delta _{ij} \hbar f(x,p)$; with $f(x,p)$ representing the effects of quantum gravity. The capital $X$ and $P$ indicate that the position and momentum operators are changed from their canonical form. A common example that we will refer to often in this work is the modified commutator of \cite{KMM} of the form
\begin{equation}
    \label{KMM-com}
    [{\hat X}_i, {\hat p}_j] = i \delta _{ij} \hbar (1 + \beta |{\vec p}|^2) ~.
\end{equation}
In this model the position and momentum as given by
\begin{equation}
    \label{KMM-op}
    {\hat X}_i = i \hbar (1 + \beta |{\vec p}|^2) \frac{\partial}{\partial p_i} ~~~{\rm and}~~~ {\hat p}_i = p_i ~,
\end{equation}
{\it i.e.} the position operator is modified but the momentum operator is not. 
The constant $\beta$ is a phenomenological parameter that characterizes the scale at which quantum gravity effects become important. Conventionally, it is thought $\beta$ should be of the Planck scale {\it i.e.} $\beta \sim \frac{l^2_{Pl}}{\hbar ^2}$ with $l_{Pl}$ being the Planck length. A full analysis of the system in equations \eqref{KMM-com} and \eqref{KMM-op} is given in reference \cite{KMM}, but for our purposes we recall two important results for this particular GUP model:
\begin{itemize}
    \item Equations \eqref{KMM-com} and \eqref{KMM-op} have a minimum length of $\Delta |{\vec x}| = \hbar \sqrt{\beta}$ at $\Delta |{\vec p}| = \frac{1}{\sqrt{\beta}}$
    \item In order for position and momentum operators to be symmetric {\it i.e.} $(\langle \psi | p_i ) | \phi \rangle = \langle \psi | (p_i  | \phi \rangle)$ and $(\langle \psi | x_i ) | \phi \rangle = \langle \psi | (x_i  | \phi \rangle)$, the scalar product of this model needs to be given by 
    \begin{equation}
        \label{scalar}
        \langle \psi | \phi \rangle = \int _{- \infty} ^{\infty} \frac{d^3p}{1+\beta |{\vec p}|^2} \psi ^* (p) \phi (p).
    \end{equation}
\end{itemize}
The modification of the scalar product as given by \eqref{scalar} is for three dimensions, but in $n$ dimensions one still has the same modifying factor for the momentum integration, namely $\frac{d^np}{1+\beta |{\vec p}|^2}$. More generally, for a modified position operator of the form 
\begin{equation}
    \label{gen-x}
    {\hat X}_i  = i \hbar f(|{\vec p}|^2) \frac{\partial}{\partial p_i} ~,
\end{equation}
the scalar product must take the form
    \begin{equation}
        \label{scalar-1}
        \langle \psi | \phi \rangle = \int _{- \infty} ^{\infty} \frac{d^np}{f(|{\vec p}|^2)} \psi ^* (p) \phi (p) ~.
    \end{equation}
These results from \eqref{scalar} and \eqref{scalar-1} will become important in the next section. 

\section{GUP and its Effects on  Vacuum Energy Calculations}

\subsection{Vacuum Energy in KMM GUP} 

The main issue we want to examine is how GUP affects the calculation of the vacuum energy and cosmological constant as laid out in \eqref{rho-cc}, \eqref{rho-vac}, and the surrounding discussion. One of the earliest and most impactful works dealing with the cosmological constant problem in the context of GUPs is the work by Chang {\it et al.} \cite{chang}. In their work, the authors calculate how the GUP, as defined by \eqref{KMM-com} and \eqref{KMM-op}, modifies Liouville's theorem and the phase space volume, {\it i.e.} $d^n x ~d^n p$, in $n$ spatial dimensions. The modified phase space found in \cite{chang} for the GUP from \eqref{KMM-com} and \eqref{KMM-op} is
\begin{equation}
    \label{phase-space-1}
    \frac{d^n x ~ d^n p}{(1+ \beta |{\vec p}|^2)^n} ~.
\end{equation}
The volume in \eqref{phase-space-1} is integrated out ($\int d^n x \to V$). Upon quantization, the claimed phase space volume from \cite{chang} becomes
\begin{equation}
    \label{phase-space-2}
    \frac{V ~ d^n p}{(2 \pi )^n (1+ \beta |{\vec p}|^2)^n} ~.
\end{equation}
Recall we are using units with $\hbar =1$ as consistent with reference \cite{weinberg}. Thus, to compare \eqref{phase-space-2} with the result in \cite{chang}, one should replace the factor $2 \pi$ by $2 \pi \hbar$ in the denominator above. 
Using the result in \eqref{phase-space-2} for three spatial dimensions, the calculation of $\rho_{vac}$ via \eqref{rho-vac} changes to
\begin{equation}
    \label{rho-vac-1}
    \rho_{vac} = \int \frac{d^3p}{(2\pi)^3 (1+ \beta |{\vec p}|^2)^3} \frac{1}{2} \sqrt{{\vec p}^2 +m^2} = \frac{1}{2} \int _0 ^\infty \frac{4 \pi}{(2 \pi)^3} dp ~ \frac{p^2 \sqrt{p^2 +m^2}}{(1+ \beta |{\vec p}|^2)^3} ~.
\end{equation}
Since the integrand of \eqref{rho-vac-1} is ${\cal O} \left( \frac{1}{|{\vec p}|^3} \right)$ at large momentum, it is convergent and does not need to have the $dp$ integration capped as in \eqref{rho-vac}. One can integrate \eqref{rho-vac-1} exactly for any $m$ \cite{chang}; for the sake of simplicity, when $m=0$, \eqref{rho-vac-1} becomes
\begin{equation}
    \label{rho-vac-2}
    \rho_{vac} (m=0) = \frac{1}{16 \pi ^2 \beta ^2}~.
\end{equation}
If one takes $\beta$ to be of the Planck scale, then the result from \eqref{rho-vac-2} still leaves the GUP modified vacuum energy to be about 118 orders of magnitude larger than the measured vacuum energy of $\rho _{vac} \approx 10^{-47}$ GeV$^{-4}$. In fact, by comparing \eqref{rho-vac} and \eqref{rho-vac-2} and using dimensional analysis, one finds that $\beta \sim p_c ^{-2}$. Thus, using the GUP cutoff factor of $\frac{1}{(1+ \beta |{\vec p}|^2)^3}$, while making $\rho_{vac}$ finite, still leaves $\rho_{vac}$ much too large which fails to resolve the cosmological constant problem. One has only replaced the ``by hand" cutoff in \eqref{rho-vac} with the functional cutoff of \eqref{rho-vac-1}. This failure of the GUP, defined by \eqref{KMM-com} and \eqref{KMM-op}, to address the cosmological constant puzzle was already noted in \cite{chang}.

However, there may be an additional problem with the integration over the momentum in \eqref{rho-vac-1}: it appears to disagree with the momentum integration from \cite{KMM}, as given by the definition of the scalar product in \eqref{scalar} or more generally in \eqref{scalar-1}. In the momentum space integration in \eqref{scalar}, there is only one factor of $(1+ \beta |{\vec p}|^2)$ in the denominator, as compared to the denominator of \eqref{rho-vac-1}, which has three factors of $(1+ \beta |{\vec p}|^2)$. If one only had one factor of $(1+ \beta |{\vec p}|^2)$ in the denominator of \eqref{rho-vac-1}, as implied by \eqref{scalar}, then the integrand would go as ${\cal O} \left( |{\vec p}| \right)$ and would thus diverge. 

The derivation of the phase space volume carried out in \cite{chang} that gave the result in \eqref{phase-space-1} is long, but straight forward, so it is hard to see any problem with this result. On the other hand, having a momentum space volume that has a factor of $(1+ \beta |{\vec p}|^2)^{-n}$ for the $d^n p$ integration would then violate the symmetry of the position operator which is the requirement that led to \eqref{scalar}; that is if the momentum integration in \eqref{rho-vac-1} is correct then this would imply $(\langle \psi | x_i ) | \phi \rangle \ne \langle \psi | (x_i  | \phi \rangle)$. 

One potential solution to the difference in the integration factors between \eqref{scalar} and \eqref{phase-space-2} could be to reconsider the spatial/volume calculation.  In the transition from \eqref{phase-space-1} to \eqref{phase-space-2}, it is assumed that the real spatial volume with GUP is the same as without GUP, that is, $\int d^n x = V$.  The introduction of a minimal length may change the calculation of volumes in some way.
If one could argue the $n-1$ factors of  $(1+ \beta |{\vec p}|^2)$ should go with the $d^nx$ integration, this would leave the correct single factor of $(1+ \beta |{\vec p}|^2)$ to go with the $d^np$ integration.  This would resolve the discrepancy between \eqref{scalar} and \eqref{phase-space-1}. Ordinarily, all the factors of  $(1+ \beta |{\vec p}|^2)$ should fall under the $d^np$ integration, but in the GUP given by \eqref{KMM-com} and \eqref{KMM-op} one can see that the position operator becomes dependent on the momentum. 
We suggest that in spherical coordinates every length $r$ should carry with it a factor of $(1+ \beta |{\vec p}|^2)^{-1}$.  
The $n$ dimensional version of the GUP modified phase space given in \eqref{phase-space-1} should be written as
\begin{equation}
    \label{phase-space-4}
    \left( \frac{d^n x}{(1+ \beta |{\vec p}|^2)^{(n-1)}} \right) \left( \frac{d^n p}{(1+ \beta |{\vec p}|^2)} \right) 
    =    \left( \frac{r^{n-1} dr d\Omega}{(1+ \beta |{\vec p}|^2)^{(n-1)}} \right) \left( \frac{d^n p}{(1+ \beta |{\vec p}|^2)} \right)~.
\end{equation}
For low energy/momentum, where $\beta |{\vec p}|^2 \ll 1$, the length will not change much, but for high energy/momentum, where $\beta |{\vec p}|^2 \gg 1$, the length is reduced.
This way, the modified momentum integration from the requirement of symmetry of the position and momentum operators as given in \eqref{scalar} and the GUP modified phase space of \eqref{phase-space-1} now agree. 

If the momentum space integration is now given by one factor of $(1+ \beta |{\vec p}|^2)$, as implied by \eqref{phase-space-4}, rather than $n$ factors as implied by \eqref{phase-space-2} or \eqref{rho-vac-1}, then not only does the GUP of equations \eqref{KMM-com} and \eqref{KMM-op} not solve the cosmological constant puzzle, as already acknowledged in \cite{chang}, but the cosmological constant is not even finite. 
In the next subsection, we will investigate a different GUP, which does give a finite cosmological constant and examine to what extent this different GUP can address the cosmological constant problem.

\subsection{Alternative GUP and the Associated Vacuum Energy}

From the generalized modified position operators of \eqref{gen-x} and the associated modified momentum integration in \eqref{scalar-1}, one can see that the integrand for $\rho_{vac}$ will be of order ${\cal O} \left( \frac{|{\vec p}|^3}{f(|{\vec p}|)} \right)$. Thus, $f(|{\vec p}|)$ must have a dependence of $|{\vec p}|^5$ or higher for the integral to be finite. One such GUP that meets this requirement is given in the recent paper \cite{BCLS} which has modified operators in three spatial dimensions of the form
\begin{equation}
    \label{tanh}
    X_i = i \hbar\cosh^2\left(\frac{|\vec{p}|}{p_M}\right)\partial_{p_i} ~~~~;~~~~ P_i = \frac{p_ip_M}{|\vec{p}|} \tanh\left(\frac{|\vec{p}|}{p_M}\right)~.
\end{equation}
From \eqref{gen-x}, we see that \eqref{tanh} implies $f(|{\vec p}|) = \cosh^2\left(\frac{|\vec{p}|}{p_M}\right)$. Thus, the GUP in \eqref{tanh} implies a vacuum energy density of 
\begin{equation}
    \label{rho-vac-4}
    \rho_{vac} ^{tanh} = \int \frac{d^3p}{(2\pi)^3 \cosh^2\left(\frac{|\vec{p}|}{p_M}\right)} \frac{1}{2} \sqrt{({\vec P})^2 +m^2} \approx \frac{1}{2} \int _0 ^\infty \frac{4 \pi}{(2 \pi)^3} dp ~ \frac{p^2 p_M \tanh \left(\frac{p}{p_M}\right) }{\cosh^2\left(\frac{p}{p_M}\right)} 
\end{equation}
which has an integrand that exponentially decays with momentum.  
In \eqref{rho-vac-4} we set the rest mass equal to zero ($m=0$), and used $|{\vec P}| = \frac{|\vec{p}|p_M}{|\vec{p}|} \tanh\left(\frac{|\vec{p}|}{p_M}\right) = p_M \tanh\left(\frac{p}{p_M}\right)$. One can evaluate the last expression exactly and this yields a finite answer 
\begin{equation}
    \label{rho-vac-5}
    \rho_{vac} ^{tanh} = \frac{p_M ^4 \ln (2)}{4 \pi ^2} ~.
\end{equation}
Thus, with this GUP model we do get a finite vacuum energy density while maintaining symmetry of the position and momentum operators.
In contrast the GUP model given by equations \eqref{KMM-com} and \eqref{KMM-op}, has an infinite vacuum energy density  when {\it only one} power of $1+\beta |{\vec p}|^2$ (as argued in this work) is used in the denominator of the vacuum energy density \eqref{rho-vac-1}.  
Even for the GUP models like that in \eqref{tanh}, where the vacuum energy density is finite, the end conclusion is essentially the same as for the vacuum energy density in \eqref{rho-vac} which is obtained via a ``by-hand" cutoff: both go as momentum to the fourth power. 
Comparing the vacuum energy densities from \eqref{rho-vac}, \eqref{rho-vac-2}, and \eqref{rho-vac-5}, they all have essentially the same form, with different notations for the momentum scale cut-off.  
Thus, whether the vacuum energy is infinite and cut-off ``by-hand" or is finite due to using a GUP like \eqref{tanh}, both these models are equally ineffective at addressing the cosmological constant problem.  

\section{Summary and conclusions}

In this work, we have examined how the GUP may alter the calculation of the vacuum energy density and the related cosmological constant. In standard QFT, which was reviewed in section I, the vacuum energy diverges and must be cut-off as in \eqref{rho-vac}, which leads to a quartic dependence of the vacuum energy density on the cut-off. 

GUP models with their associated minimal length scales provide a potential avenue to calculate a finite vacuum energy density. Having a minimal length implies a maximum energy-momentum which cuts off the divergence in the standard vacuum energy density given in \eqref{rho-vac}. An early work \cite{chang} led to a finite vacuum energy density given by \eqref{rho-vac-1} and \eqref{rho-vac-2}. However, one of our points was to argue that the calculation of the vacuum energy given in \cite{chang}  by \eqref{rho-vac-1} is inconsistent with the requirement that the position and momentum operators are symmetric in GUP models such as \cite{KMM}. This symmetry requirement leads to an integration over momentum as given in \eqref{scalar} for the GUP from equations \eqref{KMM-com} and \eqref{KMM-op} or for a more general modified position as in \eqref{scalar-1}.  Although, if one takes only a single factor of $1+ \beta |{\vec p}|^2$ in the momentum integration used to calculate $\rho _{vac}$, then one finds that the vacuum energy density from the GUP is not finite, which conflicts with the results of \cite{chang} which has $n$ factors of $1+ \beta |{\vec p}|^2$. 
Alternatively, one can preserve the symmetry of the position and momentum operators, but then the vacuum energy density is infinite for some GUPs like \eqref{KMM-com} and \eqref{KMM-op}. In the present work, we argued for the latter option, because when arriving at the momentum integration measure of \eqref{rho-vac-1}, one had to integrate out the spatial volume, as is done in going from \eqref{phase-space-1} to \eqref{phase-space-2}. However, the implication is that one is treating the spatial volume the same as in a theory with {\it no minimal length}. 
In order to take into account the minimal length of the GUP, $n-1$ of the $n$ factors of $1+ \beta |{\vec p}|^2$  should correspond to the volume integration to take into account the minimal length, leaving one factor to go with the momentum integration. 

In a larger sense, GUPs may not be able to resolve the cosmological constant problem.  
We presented an GUP model \eqref{tanh} where the integrand in the vacuum energy density decayed exponentially and led to a finite integral.  
However, this led to the same quartic momentum behavior as the in ``by-hand'' cutoff of \eqref{rho-vac} which were all essentially the same up to multiplicative factors of order one.
Regardless, the end result for all the models is more or less the same. 

There may be a way for a GUP model to address the cosmological constant problem by requiring the function $f(|{\vec p}|)$, which gives the modification for the position operator, take negative values for some range of $|{\vec p}|$.  All the GUP functions discussed here ({\it i.e.}  $1+ \beta p^2$ or $\cosh (p/p_M)$) were positive definite. In integrating the momentum from $0$ up to the QCD scale of $\Lambda _{QCD} \sim 200$ MeV, one already had a huge disagreement between the observed and theoretically calculated vacuum energy density. 
To compensate for this already large disagreement, a GUP function that is negative for some range of $|{\vec p}|$ beyond the QCD scale is needed to cancel out the positive contribution from the low momentum part of the integration. 
This is reminiscent of the supersymmetry approach to the cosmological constant problem where the positive bosonic contribution to the vacuuum energy density is canceled by the negative fermionic contribution. Note, that requiring $f(|\vec{p}|)$ to be negative is similar to a parity transformation $\vec{x} \to -\vec{x}$ but is an unusual parity transformation in that it is not discrete but rather changes continuously as momentum increases.  In any case, this may provide a fruitful new avenue for addressing the cosmological constant problem with GUPs.

\end{document}